\documentclass[adp,fleqn]{w-art}
\usepackage{times}
\usepackage{w-thm}
\usepackage[]{graphicx}
\usepackage{cite}
\begin{document}
\DOIsuffix{theDOIsuffix}
\Volume{12}
\Issue{1}
\Copyrightissue{01}
\Month{01}
\Year{2003}
\pagespan{1}{}
\Receiveddate{23 January 2009}
\Accepteddate{14 February 2009 by U.\ Eckern}
\title[]{Stripe segregation and magnetic coupling in the nickelate\\ La$_{5/3}$Sr$_{1/3}$NiO$_4$}
\author[]{Udo Schwingenschl\"ogl$^{1,2,}$\footnote{Corresponding
     author \quad E-mail: {\sf udo.schwingenschloegl@physik.uni-augsburg.de}}} 
\address[]{$^1$Institut f\"ur Physik, Universit\"at Augsburg, 86135 Augsburg, Germany}
\address[]{$^2$KAUST, PCSE Division, P.O.\ Box 55455, Jeddah 21534, Saudi Arabia}
\author[]{Cosima Schuster$^1$}
\author[]{Raymond Fr\'esard$^3$}
\address[]{$^3$Laboratoire CRISMAT, UMR CNRS-ENSICAEN(ISMRA) 6508, and IRMA, FR3095 Caen, France}
\renewcommand{\leftmark}{U.\ Schwingenschl\"ogl et al.:
Stripe segregation in La$_{5/3}$Sr$_{1/3}$NiO$_4$}

\begin{abstract}
We investigate the consequences of the stripe formation in the nickelate
La$_{5/3}$Sr$_{1/3}$NiO$_4$ for the details of its crystal structure and
electronic states. Our data are based on numerical simulations within density
functional theory (DFT) and the generalized gradient approximation (GGA). The 
on-site Coulomb interaction is included in terms of the LDA+U scheme. 
Structure optimization of preliminary experimental data indicates a strong
interaction between the structural and electronic degrees of freedom. In
particular, we find a segregation of the diagonal filled stripes induced by 
a delicate interplay with the magnetic coupling. Beyond the cooperative effect
of stripe segregation and spin order, distinct octahedral distortions are
essential for the formation of an insulating state.
\end{abstract}

\maketitle

\section{Introduction}
Doped Mott insulators present a wealth of intriguing properties with a great
potential for technological applications. They are either based on
metalicity, like high-T$_c$ superconductivity and
thermoelectricity, or trace back to an insulating behavior, like colossal
magneto-resistance and multiferroics. Therefore, understanding the origin of
the insulating state of these doped Mott insulators has clear implications, be
it caused by a lack of coherence  \cite{Kyung06} or by a formation of an
inhomogeneous state, as a stripe phase, for instance
\cite{Sac95,Tra96,Lee97,Yos00,Lee02,Fre02,Huc06}. 
Though intensely debated, see Ref.\ \cite{Kiv} for a review, there is growing
consensus that electronic models can account for the poorly metallic character of
the diagonal filled stripes observed in cuprates, almost irrespective of
their microscopic structure. For instance, magnetic diagonal filled
stripe phases of the Hubbard model are insulating \cite{Rac06a},
while a poorly metallic phase with nodal quasiparticles characterizes flux
diagonal filled stripe phases of the $t-J$ model \cite{Rac07}. In contrast,
similar model calculations devoted to isostructural nickelates fail to reproduce an
insulating groundstate \cite{Rac06b}.  

Experimentally, the stripe order in ${\rm La_{2-x} Sr_x NiO_4}$ and
${\rm Nd_{2-x} Sr_x NiO_4}$ has been evidenced by spin-polarized
as well as unpolarized neutron scattering. With decreasing temperature, the
appearance of peaks at the wave vectors ${\bf Q}_{c}=2\pi(\epsilon,\epsilon)$ 
and ${\bf Q}_{s}=\pi(1\pm\epsilon,1\pm\epsilon)$, respectively, with 
$\epsilon\simeq x$, is interpreted as onset of a diagonal filled stripe. The
onset temperature is sensitive to the doping, turning out to have a maximum
at $x=1/3$, i.e.\ when the charge wave vector matches right the spin wave vector. In
this case, specific heat \cite{Ram96}, transport \cite{Che94}, and optical
conductivity \cite{Kat96} data point to an ordering temperature of 
$T\sim 240$\,K. Recently, the stability of the stripe phase has been further
probed via non-linear transport experiments for the system
${\rm Nd_{1.67}Sr_{0.33}NiO_4}$ \cite{Huc07}. No depinning of the stripes has
been observed: instead, the non-linear transport is attributed to resistive
heating, thereby pointing towards a rather robust insulating state.

\begin{figure}[t]
\begin{center}
\includegraphics[width=0.35\textwidth,clip]{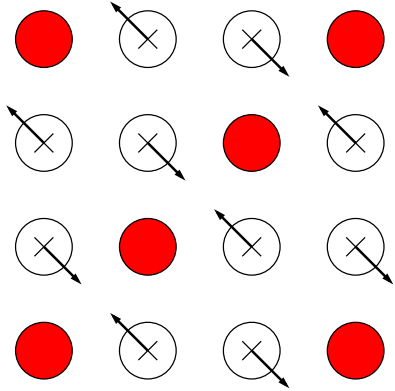}
\caption{Schematic representation of the NiO$_2$-layer in La$_{5/3}$Sr$_{1/3}$NiO$_4$
highlighting the diagonal Ni stripes. O atoms are not shown for clarity. Arrows
indicate the principal structural distortion consistent with AF spin order.}
\label{fig2}
\end{center}
\end{figure}

On the theoretical side, quite some efforts have been made to properly
describe these materials, especially by the investigation of realistic
models. While an extensive study of a two-band Hubbard model for the
e$_ g$ electrons correctly predicts diagonal filled stripes to be lowest in
energy within this class of states, the groundstate is still metallic
\cite{Rac06b}. Adding electron-lattice coupling promotes the insulating
groundstate, but the available results are limited to the $x=1/2$ case and to
particular lattice distortions \cite{Zaa94}. Band structure calculations
also have been performed \cite{Yam07,Sch08}. However, because they make use 
of an explicit lattice structure as input, which so far could not be
determined experimentally for the title compound, the validity of the
calculations could be limited. Indeed, as argued in \cite{Zaa94}, a strong
spin-charge-lattice coupling is to be anticipated in layered nickelates.
Using an two-band Hubbard model, Hotta and Dagotto \cite{hotta04}
have demonstrated the interplay of Coulombic and phononic interactions
and its relevance for the stripe formation.

Another appealing scenario of spin-lattice interaction is sketched in Fig.\ \ref{fig2},
where we show the Ni atoms for a NiO$_2$-layer. Atoms along the stripes and in the
magnetic domains are represented by full and open circles, respectively.
A Ferromagnetic (F) as well as an antiferromagnetic (AF) order between the spins
on the stripe and the domain sites is compatible with the experimental
modulation vector ${\bf Q}_{s}=\pi(2/3,2/3)$. For the AF 
alignment, the structural distortion shown in Fig.\ \ref{fig2} enhances the superexchange
over the stripes significantly, while it reduces the frustrated superexchange within the
magnetic domains only moderately. The configuration thus is expected to be subject to an
instability against a modulation of the Ni--Ni bond length perpendicular to the stripe
direction. In contrast, the F alignment in this respect is consistent with an
equal spacing of the Ni atoms, because a modulation of the bond lengths tends to
reduce the kinetic energy. The competition between the two states makes it highly
desirable to understand in detail the shape of
the lattice deformations encountered in the system, and their consequences for the
electronic structure.

\section{Technical details}
\begin{figure}[b]
\begin{center}
\includegraphics[width=0.4\textwidth,clip]{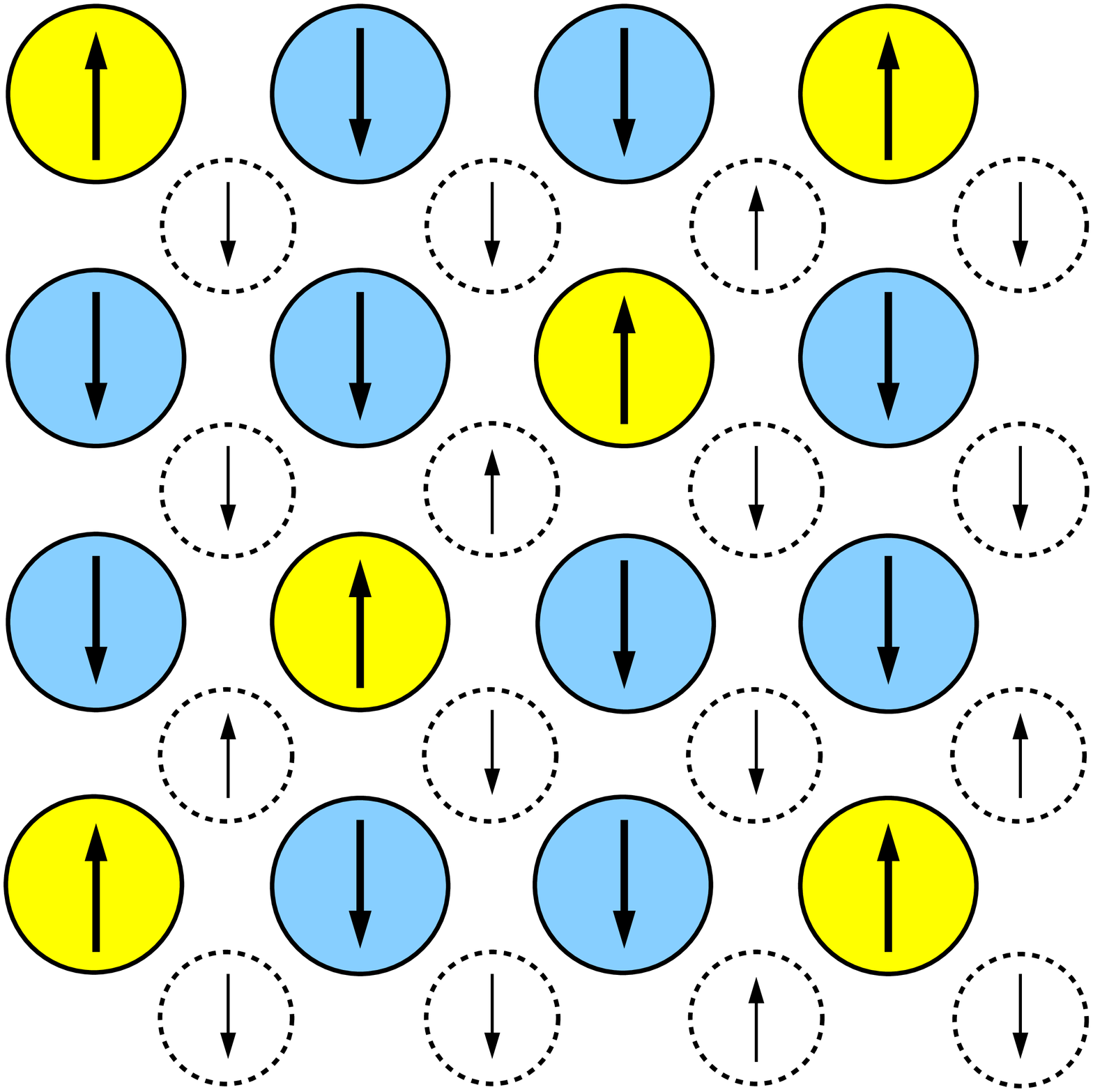}\hspace{1cm}
\includegraphics[width=0.4\textwidth,clip]{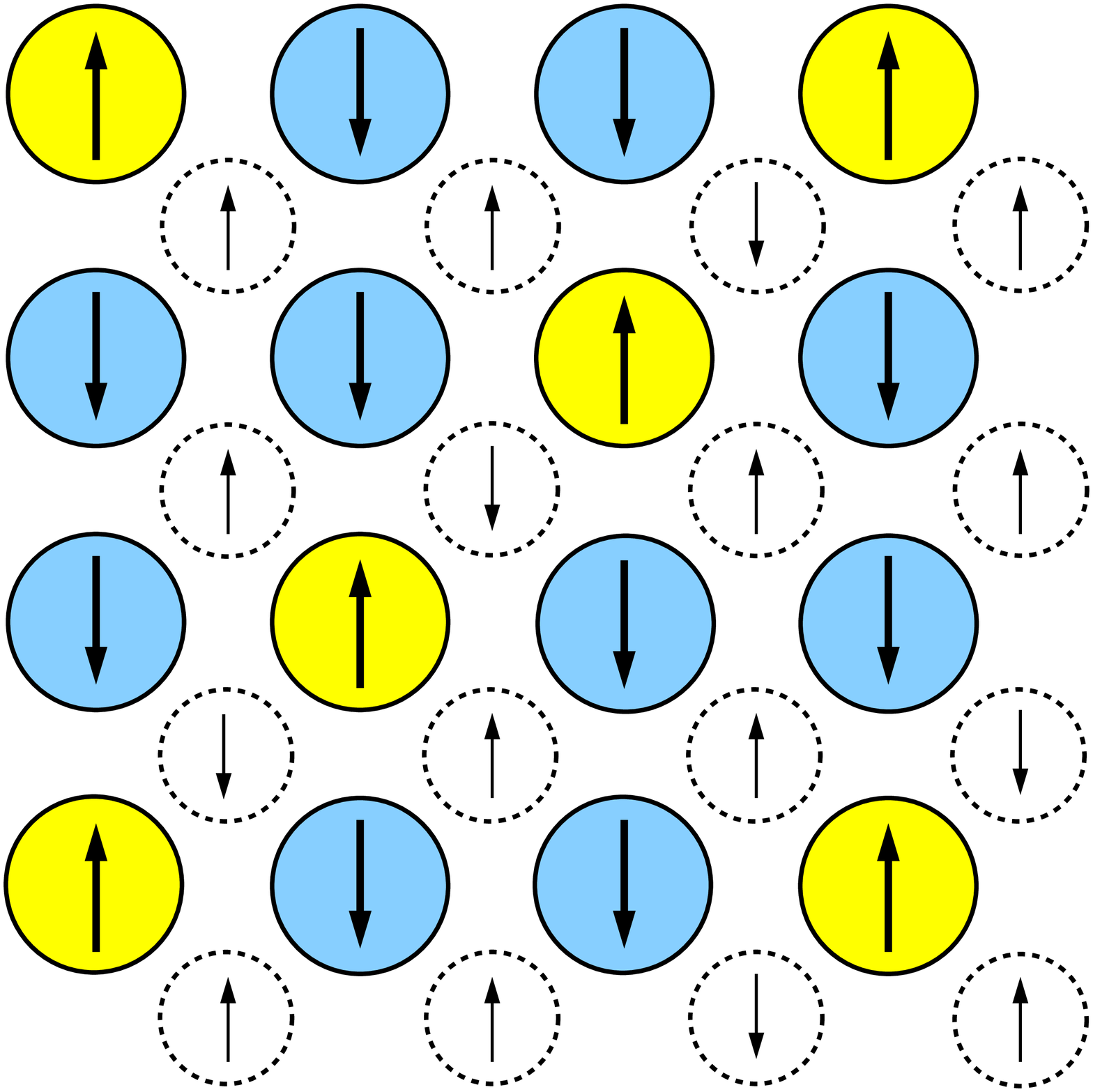}
\caption{Magnetic structure of the Ni atoms in the NiO$_2$-layers of La$_{5/3}$Sr$_{1/3}$NiO$_{4}$ for
the CDFAS (left hand side) and GDFAS (right hand side) spin configurations.
Full yellow (blue) circles denote the Ni1 (Ni3) atoms, whereas dashed circles
refer to the second NiO$_2$-layer, containing the Ni2 (Ni4) atoms, see the text.}
\label{fig_str}
\end{center}
\end{figure}

To that aim we have performed band structure calculations for the $x=1/3$
compound La$_{5/3}$Sr$_{1/3}$NiO$_4$. Our following considerations first
focus on the lattice deformations tracing back to the stripe formation, where
the parent body-centered tetragonal K$_2$NiF$_4$ structure with space group
$I4/mmm$ will act as a reference frame \cite{takeda90}. Lattice deformations
have not been reported in the literature so far, because the experimental
refinement of the crystal structure has concentrated on the lattice
parameters. In fact, refinement of the atomic positions is difficult without a
proper structure model, which, finally, is provided by our structure
optimization data. As second step, we analyze the electronic properties of the
system for different spin configurations, which allows us to determine the
magnetic groundstate of various diagonal filled antiferromagnetic stripes. The 
spin configurations under investigation are defined in table \ref{tab_def}. In
addition, the CDFAS and GDFAS patterns are illustrated in Fig.\ \ref{fig_str}.

\begin{table}
\begin{tabular}{c|c|c}
abbreviation&intraplane spin alignment&interplane spin alignment\\\hline
ADFAS & F & AF\\
CDFAS & AF & F\\
GDFAS & AF & AF
\end{tabular}
\vspace*{0.2cm}
\caption{Definitions of the A-, C-, and G-type D(iagonal) F(illed) A(ntiferromagnetic) S(tripe)
spin configurations.} 
\label{tab_def}
\end{table}

In our electronic structure calculations we use density functional theory (DFT) as implemented
in the Wien2k code \cite{wien2k}. The latter is appropriate for the present investigation, because it
allows us to perform a full structure optimization for a complex material with large unit cell
\cite{wienstruc} and to account for the electronic correlations present in the system, by the LDA+U
(local density approximation plus on-site Coulomb interaction) method \cite{wienldau}.
For the structure optimizations the generalized gradient approximation (GGA)
\cite{pbe} and for the LDA+U calculations the self-interaction correction (SIC)
\cite{anisimov93,liechtenstein95} is used.

The tetragonal ${\rm La_{2-x} Sr_x NiO_4}$ unit cell resulting from the parent
$I4/mmm$ lattice contains two formula units. It is very convenient to describe
the atomic arrangement in terms of parallel layers of NiO$_6$-octahedra,
separated by the La/Sr atoms. To be specific, we have NiO$_2$-layers parallel to
the $ab$-plane of the tetragonal unit cell, consisting of a square planar
lattice of Ni atoms, and O atoms located midway between them. Due to the
stripe charge order, we have two crystallographically inequivalent
NiO$_2$-layers comprising the metal sites Ni1/Ni3 and Ni2/Ni4, respectively. For
further details on the crystal structure, the supercell setup, as well as other technical
aspects, we refer the reader to Ref.\ \cite{Sch08}. We use the setup
described therein as the starting point of our present study. However, our new
calculations thoroughly take into account the structure relaxation induced by
the stripe charge order. To this end, we optimize the atomic positions
individually for all spin patterns under consideration.

\section{Results and discussion}
\begin{figure}[b]
\begin{center}
\includegraphics[width=0.45\textwidth,clip]{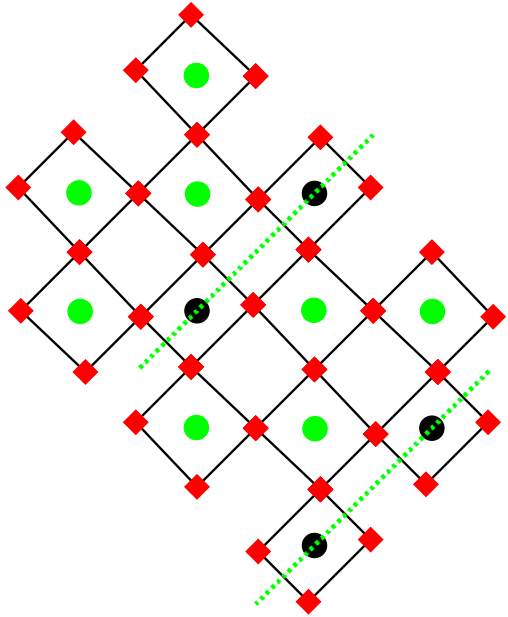}
\caption{Optimized crystal structure of a NiO$_2$-layer in La$_{5/3}$Sr$_{1/3}$NiO$_4$. The
Ni2 (Ni4) and O sites are represented by black (green) circles and diamonds, respectively,
and stripes by dotted lines.}
\label{fig_ox}
\end{center}
\end{figure}

Turning to the data obtained from our structure optimizations, we first mention that the NiO$_6$-octahedra
of La$_{5/3}$Sr$_{1/3}$NiO$_4$ are elongated and tilted with respect to the $c$-axis of the tetragonal
unit cell \cite{cava91,rod91}. While structure distortions, in general, are tiny for the ADFAS pattern,
we observe a sizeable response of the NiO$_6$-octahedra to the charge order for both the CDFAS and
GDFAS pattern, i.e.\ for AF intraplane magnetic coupling. The fact that the latter distortions are
closely related to each other supports our initial speculations about a strong spin-lattice
interaction. We summarize the observed apical and equatorial Ni--O bond lengths in the four
inequivalent NiO$_6$-octahedra in Table \ref{tab_dist}, revealing significant deviations
from the ideal values of 2.21\,\AA\ and 1.92\,\AA, respectively. The distortion associated
with these values is illustrated in Fig.\ \ref{fig_ox} for the Ni2/Ni4-plane, where the shifts
off the symmetric positions have been magnified by a factor of 4. The pattern is similar for
the Ni1/Ni3-plane, which is obtained by shifting the Ni2/Ni4-plane by the vector ($a$/2,0,$c$/2).
While for stripe ordered La$_{1.875}$Ba$_{0.125}$CuO$_4$ \cite{kim08}, for example, a
modulation of the La ions is reported, such displacements are negligible in our case for both the
La and Sr sites.

\begin{table}[b]
\begin{tabular}{c|c|c|}
atom & apical Ni--O bond length (\AA) & equatorial Ni--O bond length (\AA)\\\hline
Ni1 & 2.17 ($2\times$) & 1.94 ($4\times$)\\
Ni2 & 2.23 ($2\times$) & 1.94 ($4\times$)\\
Ni3 & 2.20 ($2\times$) & 1.89 ($2\times$) and 1.92 ($2\times$)\\
Ni4 & 2.24 ($2\times$) & 1.88 ($2\times$) and 1.94 ($2\times$)
\end{tabular}
\vspace*{0.2cm}
\caption{\rm Bond lengths in the NiO$_6$-octahedra obtained for AF intraplane magnetic coupling (CDFAS/GDFAS).}
\label{tab_dist}
\end{table}

Whereas a strong relaxation of the structure has to be
expected for a NiO$_6$-octahedron, we additionally observe a characteristic relaxation pattern
for the Ni sublattice, see the schematic illustration in
Fig.\ \ref{fig2}. With respect to the (highlighted) diagonal stripes, neighboring Ni atoms are subject
to displacements towards them. To be specific, all shifts are oriented
perpendicular to the stripe direction and parallel to the basal plane of the tetragonal unit cell, thus
parallel to the NiO$_2$-layers. Their absolute values are summarized in Table \ref{tab1}.
The distortion pattern of the Ni sublattice is well described in terms of a
segregation of neighboring stripes. By this mechanism, the magnetic intraplane
interaction is enhanced within the magnetic domains. After all, this is 
advantageous for the AF configuration only, since the F configuration is
predominantly promoted by local processes. Furthermore, our line of reasoning
is fully consistent with the fact that F intraplane coupling does not result in a
significant structure relaxation.

\begin{table}
\begin{tabular}{c|c||c|c|c}
\multicolumn{2}{c}{} &\multicolumn{3}{c}{displacement (\AA)}\\
atom moving& towards & \;ADFAS\; & \;CDFAS\; & \;GDFAS\;\\\hline
Ni3 & Ni1 chain & 0.0002 & 0.0065 & 0.0016\\
Ni4 & Ni2 chain & 0.0032 & 0.0114 & 0.0300
\end{tabular}
\vspace*{0.2cm}
\caption{\rm Atomic displacements as obtained from the structure optimizations for the different
spin patterns. The values refer to the relative shift of a Ni3/Ni4 site towards the chain formed
by the Ni1/Ni2 sites within a NiO$_2$-layer.}
\label{tab1}
\end{table}

\begin{figure}[b]
\begin{center}
\includegraphics[width=0.48\textwidth,clip]{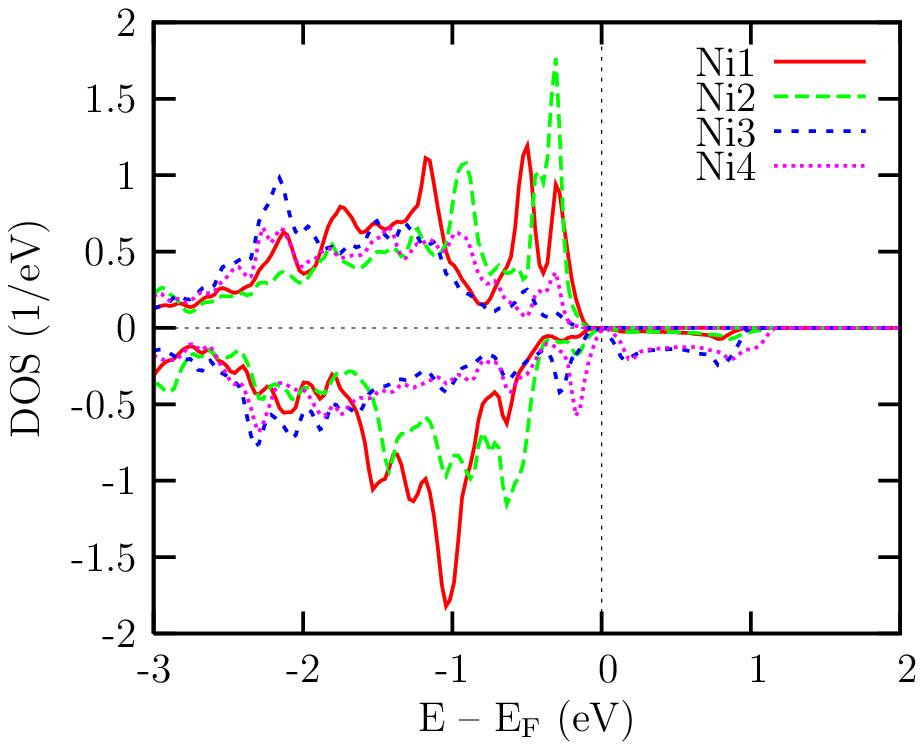}\hspace{0.3cm}
\includegraphics[width=0.48\textwidth,clip]{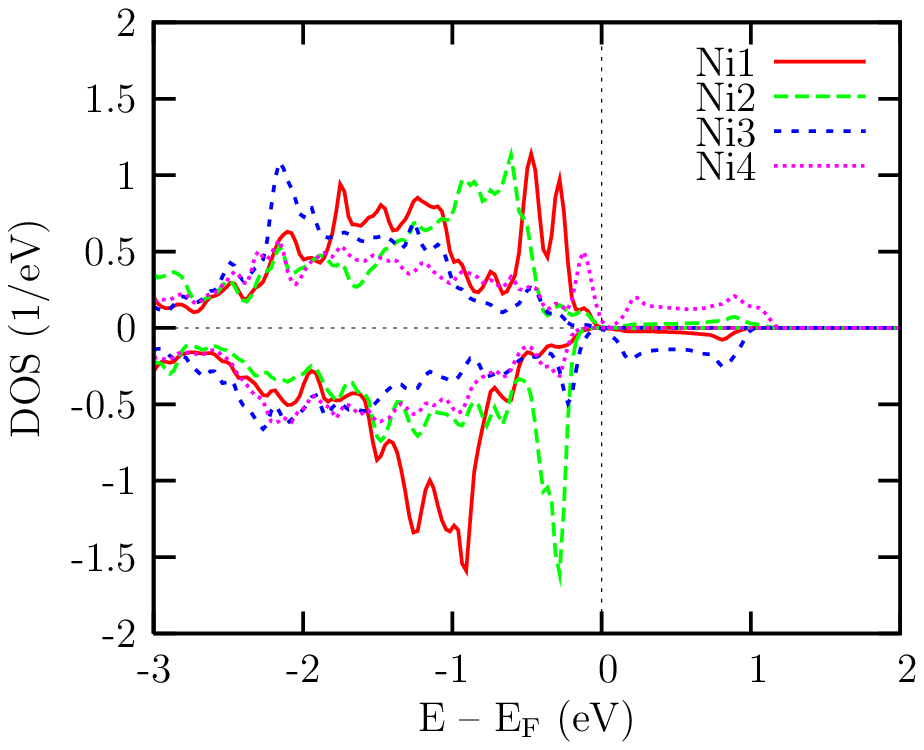}
\caption{Partial spin majority and minority Ni $3d$ DOS (per atom) for
the four inequivalent Ni sites in the CDFAS (left hand side) and GDFAS (right hand side)
spin patterns of La$_{5/3}$Sr$_{1/3}$NiO$_4$.}
\label{fig3}
\end{center}
\end{figure}

Concerning the electronic states, structural relaxation has serious effects. In contrast to
experimental findings, previous calculations for the non-optimized geometry resulted in metallic
solutions with broad conduction bands \cite{Sch08}. This behavior appeared to be characteristic
for all spin configurations involving parallel spins along the stripes, particularly for our ADFAS, CDFAS, and GDFAS
patterns. An insulating band gap is observed only for spin configurations involving antiparallel spins along the stripes \cite{Yam07}.
While the density of states (DOS) for the ADFAS phase is subject to only small changes, the
structure optimizations for the CDFAS/GDFAS phase yield substantial alterations.
The associated spin-polarized partial Ni $3d$ DOS, split up into contributions from the four inequivalent
Ni atoms, is depicted in Fig.\ \ref{fig3}. These results are based on the LDA+U scheme, with local
interactions on all Ni sites. For the on-site Coulomb interaction we set $U=8$\,eV and for the
Hund's rule coupling $J_H=0.8$\,eV \cite{anisimov91,anisimov99}.

Fig.\ \ref{fig3} seems to indicate the formation of an insulating band gap, as the consequence of a
separation of valence bands just above the Fermi energy. Closer inspection of the band structure,
however, shows that this splitting is incomplete, which could be attributed to the fact that the
on-site interaction has not been taken into account in the structure optimizations, for technical
reasons. Nevertheless, it is clear that the formation of an insulating state is intimately connected to
the relaxation of the O octahedra and the stripe segregation.

For the optimized crystal structures, we now compare the spin patterns under consideration from the
energetic point of view. The energy gain (with respect to the paramagnetic
energy calculated for the non-optimized La$_{5/3}$Sr$_{1/3}$NiO$_4$ structure) of the magnetic solutions is summarized in Table \ref{tab2}. 
In addition to the solutions displayed in Fig.\ \ref{fig3}, it seems to be possible to obtain metallic solutions with similar
energies. However, they are more indicative of a poor convergence than a physically relevant result.
All magnetic phases are lower in energy than the paramagnetic phase, already for $U=0$. In this
case, the largest total energy gain amounts to 19.5\,mRyd per Ni atom for the CDFAS phase. However,
on inclusion of the on-site Coulomb interaction the energy order is inverted, which contradicts the
experimental situation. The ADFAS pattern now is favored by 6\,mRyd per Ni atom, while the CDFAS
and GDFAS configurations are almost degenerate. Since the energy differences of the various phases are
rather small, this fact again is explained by the interplay between the magnetic
ordering and details of the crystal structure. Probably, a structure optimization within
the LDA+U scheme is required to restore the original energy order.

\begin{table}
\begin{tabular}{c|c||c|c|c}
\multicolumn{2}{c}{} &\multicolumn{3}{c}{total energy gain (mRyd)}\\
$U$ (eV) & $J_H$ (eV)& \;ADFAS\; & \;CDFAS\; & \;GDFAS\;\\\hline
0 & 0 & 8.7 & 19.5 & 17.3\\
8 & 0.8 & 163 & 156 & 157
\end{tabular}
\vspace*{0.2cm}
\caption{\rm Total energy gain per Ni atom resulting from the spin-polarization and the
simultaneous structural distortion. Values are compared for GGA and LDA+U calculations,
and for the ADFAS, CDFAS, and GDFAS spin patterns.}
\label{tab2}
\end{table}

\section{Conclusions}
In conclusion, we have presented DFT calculations for the nickelate
La$_{5/3}$Sr$_{1/3}$NiO$_4$. Optimization of the crystal structure results in
two characteristic types of lattice distortions: (1) Distinct deformations of
the NiO$_6$ octahedra pave the way for the opening of an insulating gap, which, so
far, could not be reproduced assuming parallel spins along the stripes. (2) The
strong spin-lattice interaction induces a segregation of neighboring diagonal
filled stripes, as the deformation cooperates with the AF interaction
perpendicular to the stripes. This new mechanism of stripe segregation is
expected to be relevant for a large class of striped materials with related
crystal structures. Moreover, neither the distortions of the O nor of the Ni
sublattice further reduce the symmetry initially imposed by the charge
pattern. Stripe ordered La$_{5/3}$Sr$_{1/3}$NiO$_4$ thus maintains a $Cmmm$
lattice symmetry, which to verify is an important task for future 
experiments. Besides, there is a clear need for structure optimization on the
LDA+U level, to enable a more reliable discussion of the stability of the various
phases. Work along these lines is in progress.

\begin{acknowledgement}
We acknowledge fruitful discussions with T.\ Kopp, M.\ Raczkowski, and Y.\ Sidis.
Financial support was provided by the Deutsche Forschungsgemeinschaft (SFB
484) and the Bayerisch-Franz\"osische Hochschulzentrum. 
\end{acknowledgement}

\end{document}